\documentclass[prl,10 pt,twocolumn,nofootinbib,superscriptaddress]{revtex4-1}
\usepackage{amsmath,amssymb,graphicx,color,xfrac}
\pdfoutput=1
\usepackage[mathletters]{ucs}
\usepackage[utf8x]{inputenc}

\definecolor{lightblue}{rgb}{0.17,0.39,1}
\definecolor{lightgreen}{rgb}{0.67,0.81,0.08}
\definecolor{lightred}{rgb}{1,0.05,0.52}

\usepackage[hyperfootnotes=false, bookmarks, colorlinks=true, breaklinks, %
pdftitle={Kim}, pdfauthor={Kim}]{hyperref}
\hypersetup{linkcolor=lightblue,citecolor=lightblue,filecolor=black,urlcolor=lightblue}

\newcommand{\der}[2]{\left(∂{#1}/∂{#2}\right)}

\newcommand{\Torque}{ τ }

\begin{document}
\title{Resonant torsion magnetometry in anisotropic quantum materials}

\author{K. A. Modic}
\email[Email: ]{modic@cpfs.mpg.de}
\affiliation{Max-Planck-Institute for Chemical Physics of Solids, Noethnitzer Strasse 40, D-01187, Dresden, Germany}
\author{Maja D. Bachmann}
\affiliation{Max-Planck-Institute for Chemical Physics of Solids, Noethnitzer Strasse 40, D-01187, Dresden, Germany}
\author{B. J. Ramshaw}
\affiliation{Laboratory for Atomic and Solid State Physics, Cornell University, Ithaca, NY 14853, USA}
\author{F. Arnold}
\affiliation{Max-Planck-Institute for Chemical Physics of Solids, Noethnitzer Strasse 40, D-01187, Dresden, Germany}
\author{K. R. Shirer}
\affiliation{Max-Planck-Institute for Chemical Physics of Solids, Noethnitzer Strasse 40, D-01187, Dresden, Germany}
\author{Amelia Estry}
\affiliation{Max-Planck-Institute for Chemical Physics of Solids, Noethnitzer Strasse 40, D-01187, Dresden, Germany}
\author{J. B. Betts}
\affiliation{Los Alamos National Laboratory, Los Alamos, NM 87545, USA}
\author{Nirmal J. Ghimire}
\affiliation{Los Alamos National Laboratory, Los Alamos, NM 87545, USA}
\affiliation{Argonne National Laboratory, Lemont, IL 60439, USA}
\altaffiliation[Current address]{}
\author{E. D. Bauer}
\affiliation{Los Alamos National Laboratory, Los Alamos, NM 87545, USA}
\author{Marcus Schmidt}
\affiliation{Max-Planck-Institute for Chemical Physics of Solids, Noethnitzer Strasse 40, D-01187, Dresden, Germany}
\author{Michael Baenitz}
\affiliation{Max-Planck-Institute for Chemical Physics of Solids, Noethnitzer Strasse 40, D-01187, Dresden, Germany}
\author{E. Svanidze}
\affiliation{Max-Planck-Institute for Chemical Physics of Solids, Noethnitzer Strasse 40, D-01187, Dresden, Germany}
\author{Ross D. McDonald}
\affiliation{Los Alamos National Laboratory, Los Alamos, NM 87545, USA}
\author{Arkady Shekhter}
\affiliation{National High Magnetic Field Laboratory, Florida State University, Tallahassee, FL 32310, USA}
\author{Philip J. W. Moll}
\email[Email: ]{philip.moll@cpfs.mpg.de}
\affiliation{Max-Planck-Institute for Chemical Physics of Solids, Noethnitzer Strasse 40, D-01187, Dresden, Germany}
\affiliation{EPFL STI IMX-GE MXC 240 CH-1015, Lausanne, Switzerland}

\begin{abstract}

Unusual behavior of quantum materials commonly arises from their effective low-dimensional physics, which reflects the underlying anisotropy in the spin and charge degrees of freedom. Torque magnetometry is a highly sensitive technique to directly quantify the anisotropy in quantum materials, such as the layered high-T$_c$ superconductors, anisotropic quantum spin-liquids, and the surface states of topological insulators. Here we introduce the magnetotropic coefficient $k=\partial^2 F/\partial \theta^2$, the second derivative of the free energy F with respect to the angle $\theta$ between the sample and the applied magnetic field, and report a simple and effective method to experimentally detect it. A sub-$\mu$g crystallite is placed at the tip of a commercially available atomic force microscopy cantilever, and we show that $k$ can be quantitatively inferred from a shift in the resonant frequency under magnetic field. While related to the magnetic torque $\tau=\partial F/\partial \theta$, $k$ takes the role of torque susceptibility, and thus provides distinct insights into anisotropic materials akin to the difference between magnetization and magnetic susceptibility. The thermodynamic coefficient $k$ is discontinuous at second-order phase transitions and subject to Ehrenfest relations with the specific heat and magnetic susceptibility. We apply this simple yet quantitative method on the exemplary cases of the Weyl-semimetal NbP and the spin-liquid candidate RuCl$_3$, yet it is broadly applicable in quantum materials research.

\end{abstract}

\pacs{...}
\date{\today }
\maketitle


Correlated quantum materials governed by strong electronic interactions commonly host a variety of competing and coexisting electronic phases, such as the copper- and iron-based high-T$_c$ superconductors where charge ordering, high-temperature superconductivity and magnetism occur in close proximity \cite{armitage}. Mapping the associated phase diagram is a critical first step to understanding their physics. These phases are commonly characterized by anisotropic behavior that reflects the microscopic anisotropy in the spin and charge degrees of freedom. Prominent examples include anisotropy in the magnetic susceptibility of the cuprates \cite{Ong, LuLi, ybco}, the identification of hidden-order phases in URu$_2$Si$_2$ and SmB$_6$ \cite{urusi, Lu} and the electronic nematicity of the iron-based superconductors \cite{sprau,fernandes}.

While anisotropy is at the heart of quantum materials, its experimental signatures can be very subtle. An established and highly sensitive technique to $\it directly$ probe small anisotropies in correlated metals and exotic magnets is torque magnetometry. When a sample with an anisotropic magnetization $M$ is placed in an external magnetic field $H$, it experiences a torque $\tau =M \times H$. This torque can be measured with high accuracy by mounting a crystal onto a cantilever \cite{Rossel, brugger, RSI, frank, frank2, frankbook}.

\begin{figure}[htbp]
\centering
\includegraphics[width=1.01\linewidth, trim=1.3cm 0cm 12.2cm 0cm, clip=true]{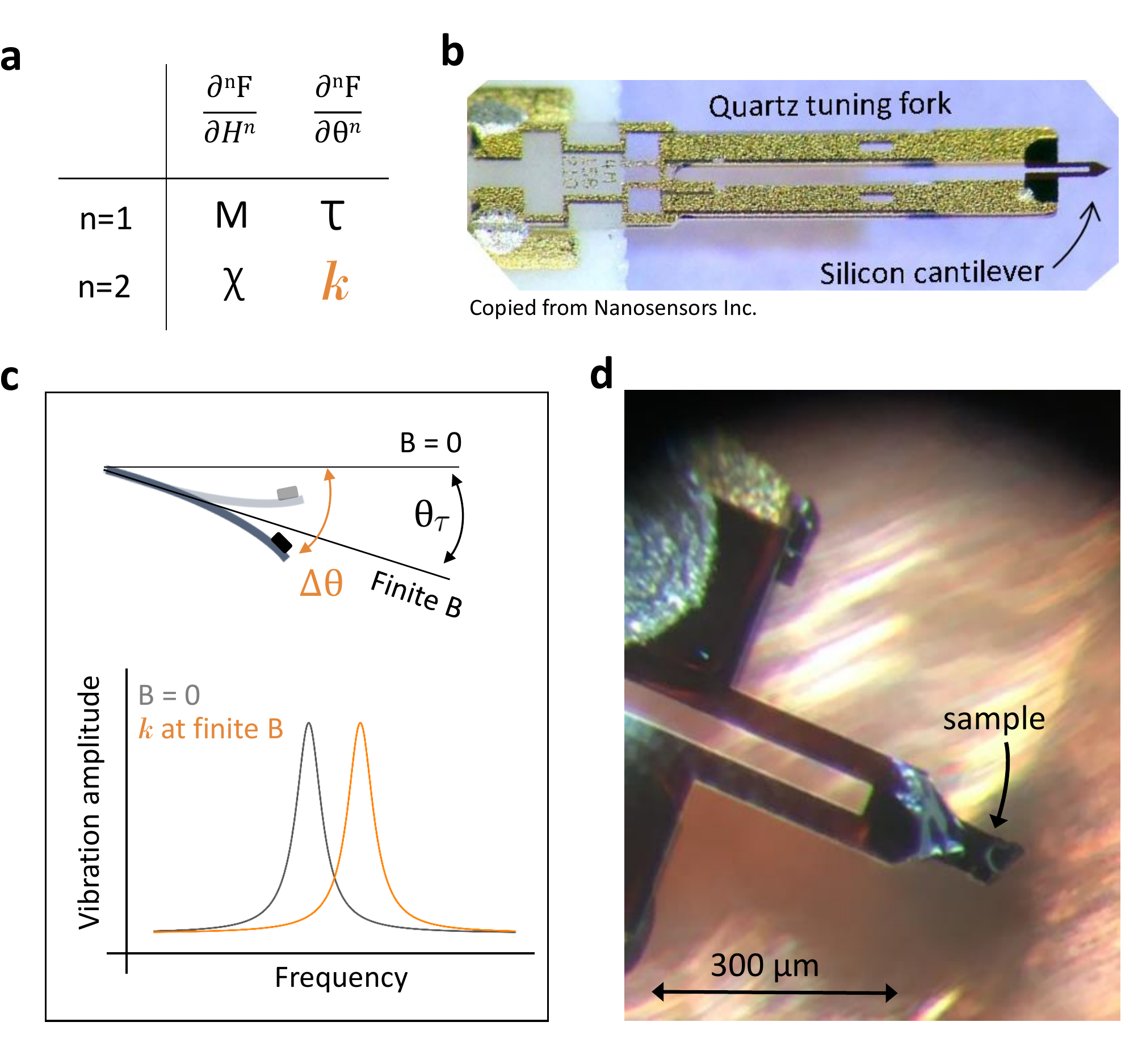}
\rule{27em}{0.5pt}
\caption[Schematic overview]{\scriptsize (a) First and second derivatives of the free energy with respect to the magnetic field $H$ and the field orientation $\theta$ (b) The quartz tuning fork of the Akiyama A-probe \cite{akiyama} is electrically excited at the lowest resonance mode of the silicon cantilever, producing a large out-of-plane motion at the tip of the cantilever. (c) Schematic representing the principle of measuring the magnetotropic coefficient $k$. In a magnetic field, the magnetic torque brings the lever to a new equilibrium position. The magnetic energy of the samples changes the effective stiffness of the lever, leading to a shift in the resonant frequency. (d) The silicon cantilever glued to each leg of the quartz tuning fork. A single crystal of RuCl$_3$ is mounted at the tip of the cantilever with Bayer silicone grease.}
\label{fig:overview}
\end{figure}

Both the magnetic torque $\tau=-\partial F/\partial θ$ and the magnetization $M=-\partial F/\partial H$ are first derivatives of the free energy $F$, and thus these thermodynamic parameters provide sensitive and essential information at phase transitions (Figure \ref{fig:overview}a). However, often the second derivatives of the free energy, such as the heat capacity $C = -T\partial ^2F/ \partial T^2$, the magnetic susceptibility $\chi = \partial ^2F/ \partial H^2$ and the elastic moduli $c_{ijkl}=\partial^2F/\partial\epsilon_{ij}\partial\epsilon_{kl}$ provide more fundamental insights into a material. They can be directly related to physical properties, such as the density of states, and are the essential quantities to formulate a microscopic theory. Unlike first derivatives, they exhibit discontinuities at second-order phase thansitions and their magnitudes can be related to one another through the Ehrenfest relations \cite{Callen}.

Here, we develop the theoretical framework of the magnetotropic coefficient $k = \partial ^2F/ \partial \theta^2$ as a second-order derivative. Related to the magnetic torque, the thermodynamic coefficient of torque susceptibility is directly linked to the magnetic anisotropy. We furthermore show a simple and effective technique for its direct measurement, ``resonant torsion magnetometry". In this approach, the sample is mounted onto a commercially-available self-resonating cantilever, and subjected to a magnetic field. While the magnetic torque itself bends the cantilever into a new static equilibrium, the magnetotropic coefficient induces an effective spring constant of the oscillator, which manifests itself into a frequency shift (Figure \ref{fig:overview}c). This approach to detect the magnetotropic coefficient is related to, yet distinct from, previous successful applications of resonant methods to detect the magnetic torque, such as ``torque differential magnetometry" on nanometer-sized samples \cite{nanotube, rings, ruthenate, allelectric, theory, nanorod, Rugar}. We demonstrate sensitivity of our probe and highlight its thermodynamic character on two exemplary quantum materials, one with charge- and the other with spin-dominated anisotropy. We measure quantum oscillations in the Weyl semimetal NbP \cite{Klotz, Shekhar2015} and the antiferromagnetic phase boundary of the spin-liquid candidate RuCl$_3$ \cite{spinliquid, spinliquid2, Banerjee, cao, johnson, heatcapacity, Baenitz}.

\color{black}
To elucidate the physical distinction between the magnetic torque and the magnetotropic coefficient and to describe the measurement, we briefly review the energetics of the resonating sample. In the harmonic approximation, the energy of a cantilever with effective stiffness $K$, moment of inertia $I$ (see methods) and an attached sample can be written as
\begin{align}\label{eq:one}
 E = \frac{I}2 \left(\frac{dΔθ}{dt}\right)^2
 +\frac{K}{2}\left({Δθ}\right)^2 -\Torque Δθ  +  \frac{k}{2}\left({Δθ}\right)^2.
 \end{align}
The first two terms describe the kinetic and potential energies of the bare cantilever and together determine the base oscillation frequency, $\omega_0^2=K/I$. We parameterize the motion of the lever as it vibrates by an angle $Δθ$ at the tip of the lever where the sample is mounted (Figure \ref{fig:overview}c). The last two terms in Eq. \ref{eq:one} describe the anisotropic energy of the measured sample in the applied magnetic field. Both the torque and the magnetotropic coefficient appear as coefficients in a Taylor expansion of the free energy $F(θ,H)$, and they manifest themselves in distinct physical responses of the sample. The torque shifts the equilibrium angle about which the lever oscillates to $θ_{\tau} =2\tau/K$ (Figure \ref{fig:overview}c). The magnetotropic coefficient encodes the curvature of the free energy with respect to the rotation angle, and appears as a shift in the oscillation frequency, $(ω_0+Δω)^2 = (K+k)/I$. For small frequency shifts, this can be expanded as
\begin{align}
\frac{Δω(θ,H)}{ω_0}  = \frac{k(θ,H)}{2K}.
\end{align}
Therefore $k$ can be directly determined by a simple measurement of the resonance frequency of the cantilever.

To illustrate the different behaviors of $\tau$ and $k$, we turn to the simple case of the linear response regime. Here, the magnetization $M$ is proportional to the magnetic field strength $H$, $M_i=\chi_{ij}H_j$ and the free energy is $F(θ, H) \propto \cos{2θ}$. Accordingly, the angle dependences of the torque $\tau \propto sin(2\theta)$ and the magnetotropic coefficient $k \propto cos(2\theta)$ strongly differ.

\begin{figure}[htbp]
\centering
\includegraphics[width=1.0\linewidth, trim=3cm 0.2cm 2.2cm 0.3cm, clip=true]{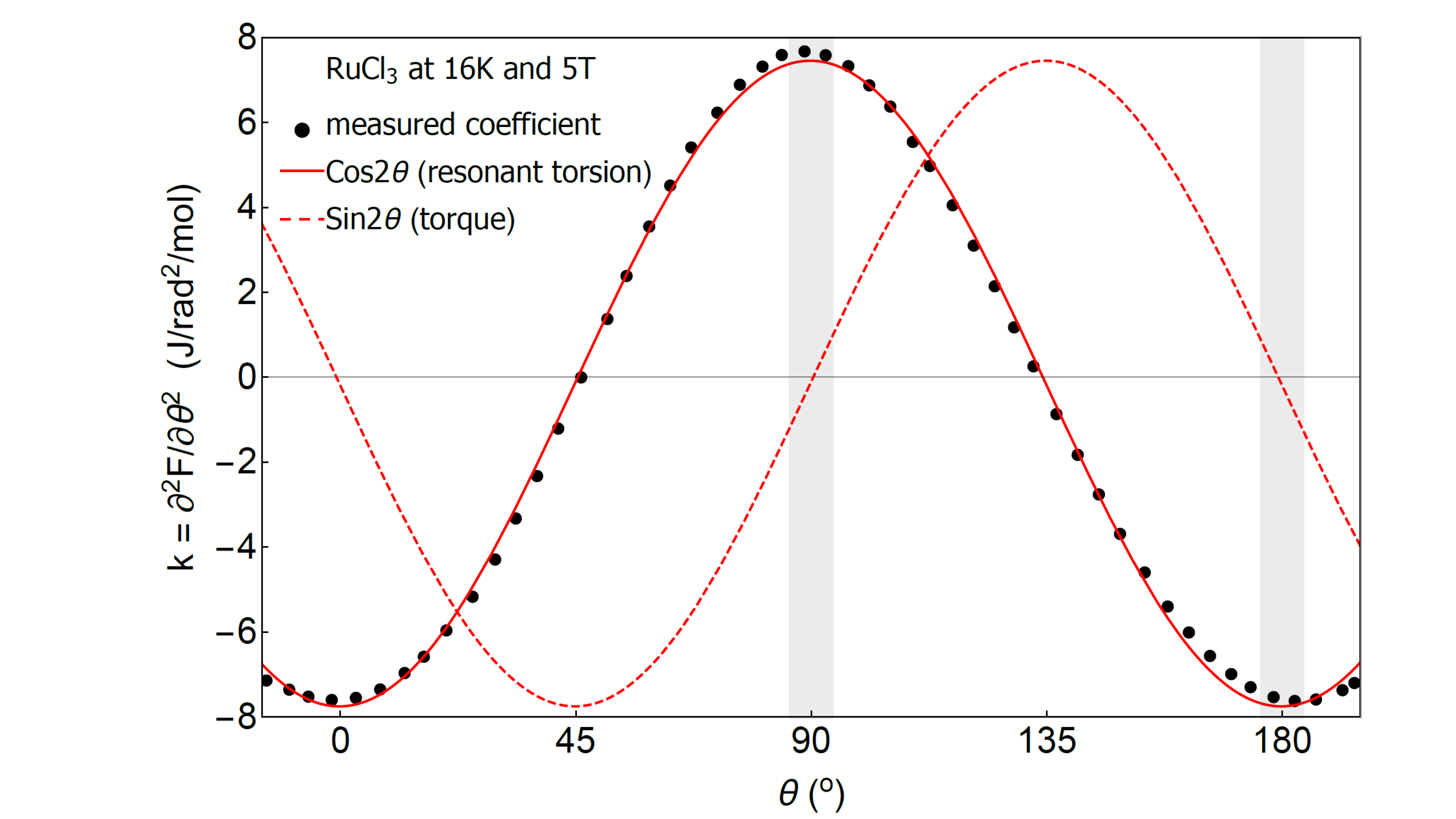}
\rule{27em}{0.5pt}
\caption[Angle Dependence]{\scriptsize The magnetotropic coefficient $k$, proportional to the shift in frequency, of RuCl$_3$ at T = 16 K and H = 5 T shows the $\cos{2θ}$ angle dependence expected in the linear resonse regime $M_i=\chi_{ij}H_j$. $\theta = 0^\circ$ and $\theta = 90^\circ$ correspond to magnetic field applied perpendicular and parallel to the honeycomb planes, respectively. The principal magnetic axes (gray bands) have a minimum and maximum response in the measured coefficient.}
\label{fig:cos}
\end{figure}

Indeed, the angle dependence expected for the magnetotropic coefficient instead of the magnetic torque is observed in a resonant torsion measurement of RuCl$_3$ at low fields within the linear regime (Figure \ref{fig:cos}). The signal of resonant torsion is thus maximal for fields along the axes of symmetry, a disadvantageous field orientation for conventional torque measurements. Along these important symmetry directions, the magnetic torque is small and subject to an undesirable torque interaction effect -- where small changes in $\theta$ lead to large changes in the magnetic torque. We note that only in the linear regime is the shift in frequency simply related to the average deflection angle. In general anisotropic quantum materials and in fields exceeding the linear response, the former is not caused by the latter and the two are independent characteristics of the magnetic anisotropy of the material.

Unlike many sensitive magnetic measurement methods, resonant torsion does not require large specialized laboratory instrumentation. Self-resonating oscillatory force sensors for scanning probe microscopy are commercially available at high quality, such as the Akiyama Probe (A-Probe) used in these experiments \cite{akiyama}. The A-probe is made of two separate resonators: a silicon U-shaped cantilever (310 $\mu$m long and 3.7 $\mu$m thick) and a quartz tuning fork (2.4 mm long and 100 $\mu$m thick) (Figure \ref{fig:overview}b) \cite{akiyama}. The lowest vibration mode of the coupled cantilever system is roughly 50 kHz and produces an out-of-plane motion at the tip of the cantilever. The frequency of this mode is highly sensitive to any external influence, which is the basis of resonant torsion microscopy.

\begin{figure}[htbp]
\centering
\includegraphics[width=1.01\linewidth, trim=0cm 0cm 0cm 0cm, clip=true]{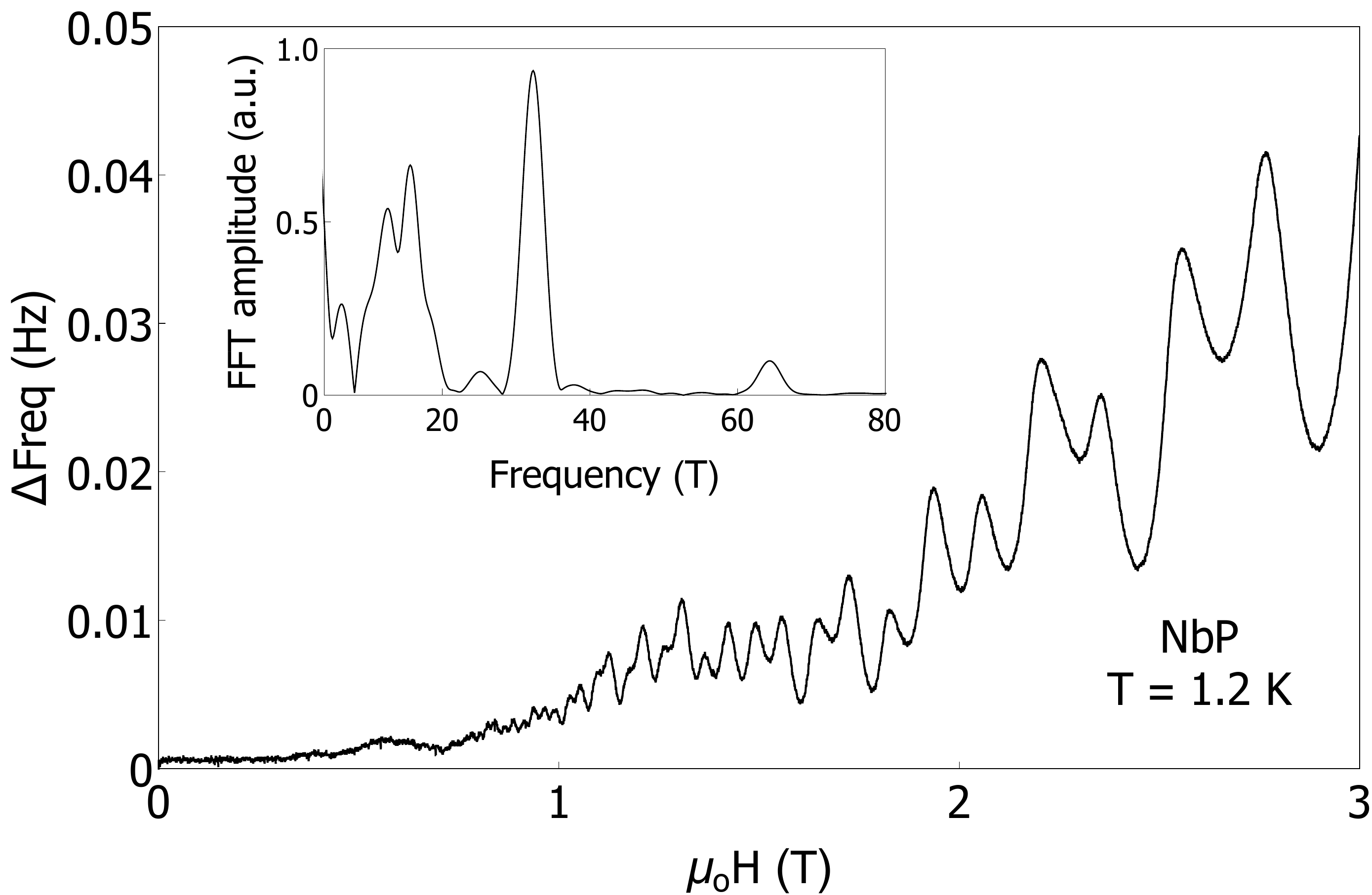}
\rule{27em}{0.5pt}
\caption[Circuit diagram]{\scriptsize At a temperature of 1.2 K, de Haas-van Alphen oscillations of NbP up to 3 T as inferred from the measured shift in frequency. Magnetic field is applied close to the crystallographic $\emph c$-axis. The measured noise with the PLL bandwidth of 1 Hz in zero magnetic field is roughly $\Delta$ = 300 $\mu$Hz.}
\label{fig:NbP}
\end{figure}

The resonance frequency as a function of temperature and magnetic field can be conveniently tracked by a phase-locked loop (see methods). In order to demonstrate the sensitivity of the technique, we measure quantum oscillations in the Weyl semimetal NbP \cite{Klotz, Shekhar2015} up to 3 T (Figure \ref{fig:NbP}). This semimetal is non-magnetic, and its entire magnetic response at low fields is due to the weak Landau diamagnetism of the conduction electrons. With the magnetic field applied along the crystallographic $\emph c$-axis, where the magnetic torque is zero, we can resolve quantum oscillations in fields well below 1 T. The quantum oscillation frequencies for this field orientation agree with those reported in the literature \cite{Klotz, Shekhar2015}. With a characteristic response bandwidth of 1 Hz, the smallest detectable frequency shift is $\Delta f/ f = 6\times10^{-9}=\Delta k/ K$, where $K$ is the effective bending stiffness of the lever (see methods). With $K$ = 180 nJ/rad$^2$, the smallest detectable magnetotropic coefficient is $\Delta k$ = $1.1\times10^{-15}$ J/rad$^2$,  equivalent to $1.2\times10^{8}$ $\mu_\text{B}$ at 1 T. This can be used to estimate the required mass of a metallic crystallite to be investigated by resonant torsion magnetometry. Even in only weakly anisotropic metals (1$\%$ anisotropy), which would contribute 0.01 $\mu_\text{B}$ per formula unit, only 10$^{12}$ formula units are needed to resolve a signal at the demonstrated sensitivity. For a 3 angstrom unit cell size, this corresponds to a 3 cubic $\mu$m sample size or a sample weight of 0.1 ng for a sample density of 5 g/cm$^3$. Resonant torsion magnetometry is thus ideally suited to investigate challenging materials where only the smallest particles exist in single crystal form.

In addition to the high sensitivity, the magnetotropic coefficient can provide valuable insight into the thermodynamics of a system via the Ehrenfest relation. $k$ can be more formally defined as a member of a matrix of second derviatives of the free energy when temperature $T$, volume $V$, magnetic field strength $|H|$, and magnetic field orientation $θ$ are independent variables. The relation of $k$ to other thermodynamic coefficients is directly apparent from the behavior of the thermodynamic potential in the $T, V, |H|$, and $θ$ variables
\begin{align}\label{eq:aaa}
dF =& -SdT  -PdV - MdH - \tau dθ.
\end{align}
In polar coordinates, $M$ and $H$ denote absolute values.

We can derive the Ehrenfest relation that relates a discontinuous jump in the resonant torsion to other thermodynamic coefficients. If we assume that $T_c(θ)$ is the boundary of a second-order phase transition induced by the magnetic field angle measured at a fixed volume $V$ and magnetic field $H$, then continuity of all first derivatives ($S, P, M, \Torque$) across such a boundary, $Δ{S} = 0$ and $Δ{\Torque} = 0$, requires that discontinuous jumps in the three thermodynamic coefficients $C$, $\der{S}{θ}=-\der{\Torque}{T}$, and $k$~are all related to each other:
\begin{align}
\frac{ΔC}{T_c} dT^* + Δ\der{S}{θ} dθ^* =0 \notag\\
 Δ\der{\Torque}{T} dT^* + Δk dθ^* =0.
\end{align}
Here  $dT^*$ and  $dθ^*$ are short segments along the phase boundary in the $T-θ$ phase plane, such that $dT^*/dθ^* = \der{T_c}{θ}$
The Ehrenfest relation connecting the jump in the magnetotropic coefficient $\Delta k$ and the jump in the heat capacity $\Delta C$ is
\begin{align}\label{eq:ehrenfest}
	Δk = -\frac{ΔC}{T_c} {\der{T_c}{θ}}_H^2,
\end{align}
where the derivative is to be taken along the phase boundary at fixed magnetic field.
Similarly, Ehrenfest relations between the jumps in $k$, $\chi$, and $C$ give $Δk = -Δ\chi \der{H_c}{θ}_T^2$ and $Δ\chi=(ΔC/T_c) \der{T_c}{H}_\theta^2$, where the derivatives in the two relations must be taken along the phase boundary at fixed temperature and at a fixed field orientation, respectively.

In order to demonstrate these thermodynamic relations, we refer again to RuCl$_3$, an effective spin-1/2 quantum magnet that orders antiferromagnetically at $T_N$ = 7 K \cite{cao}. Below this temperature, long-range order can be suppressed with a magnetic field of $\sim$8 T for fields applied within the honeycomb planes \cite{heatcapacity, Baenitz}, with recent evidence suggesting a spin liquid state at higher magnetic fields \cite{spinliquid, spinliquid2}.
\begin{figure}[htbp]
\centering
\includegraphics[width=1.01\linewidth, trim=2cm 0cm 14cm 0.5cm, clip=true]{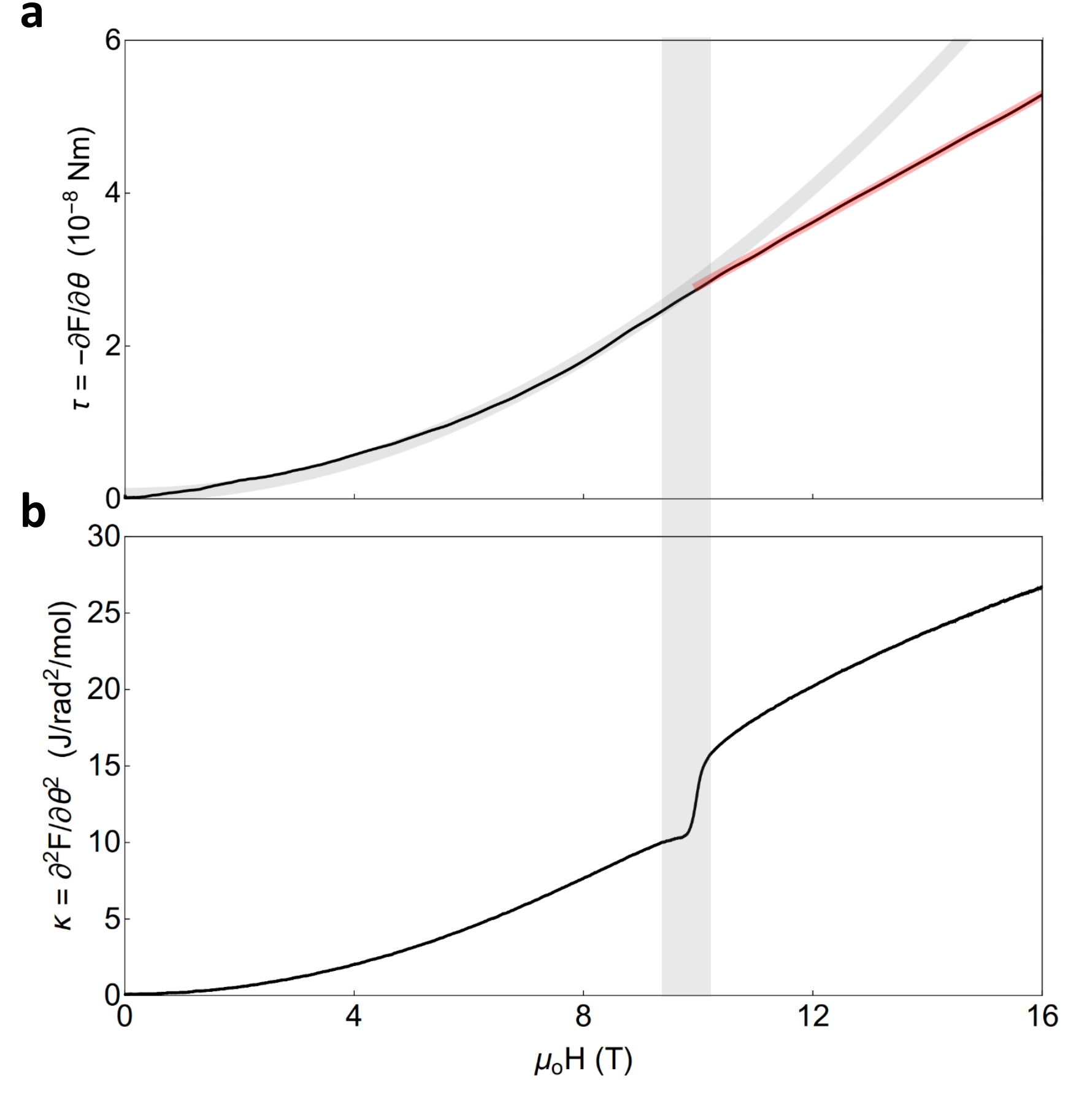}
\rule{27em}{0.5pt}
\caption[Phase transition]{\scriptsize (a) The magnetic torque $\tau$ and (b) the magnetotropic coefficient $k$ of RuCl$_3$ at T = 1.3 K. Magnetic field is applied at an angle $\sim$10$^\circ$ from the honeycomb planes toward the crystallographic $\emph c$-axis. Both measurements are set up to probe the magnetic anisotropy between the in-plane and out-of-plane field orientations (ie. $\alpha=±(\chi_\parallel-\chi_\perp)$, where $\chi_\parallel$ and $\chi_\perp$ are with respect to the honeycomb planes).}
\label{fig:jump}
\end{figure}
We measure RuCl$_3$ at T = 1.3 K -- well within the antiferromagnetically ordered state \cite{johnson, Baenitz} -- to observe the evolution of the magnetic torque and the resonant torsion as we cross the second-order phase boundary with increasing magnetic field (Figure \ref{fig:jump}). With small fields at an angle $\sim$10$^\circ$ away from the honeycomb planes, both $\tau$ and $k$ respond quadratically to the applied magnetic field. For this field orientation, we observe the suppression of long-range order at $\sim$9 T. Across the phase boundary, $\tau$ shows a break in slope crossing over to linear behavior at higher magnetic fields, whereas $k$ experiences a discontinuous jump. Akin to the advantages of techniques sensing the magnetic susceptibility compared to magnetization, detecting $k$ offers a more appropriate means for identifying magnetic phase transitions.

\begin{figure}[htbp]
\centering
\includegraphics[width=1.0\linewidth, trim=0.5cm 0cm 19cm 0cm, clip=true]{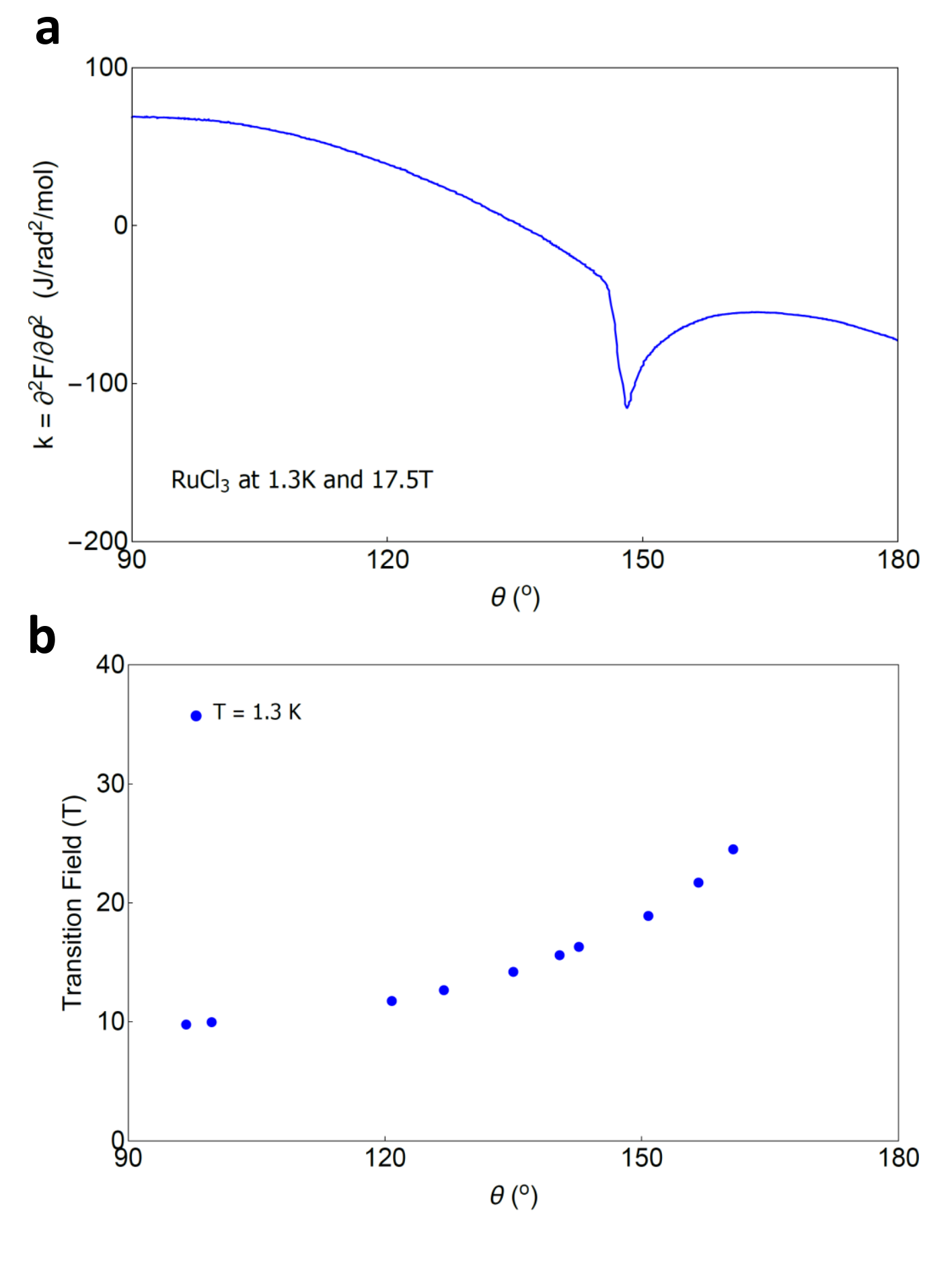}
\rule{27em}{0.5pt}
\caption[Ehrenfest]{\scriptsize (a) The magnetotropic coefficient $k$ of RuCl$_3$ at T = 1.3 K and H = 17.5 T illustrates two points of entry into the antiferromagnetically ordered state, marked by a sharp transition down (blue dots). (b) The antiferromagnetic phase boundary as determined from angle scans at finite fields. As the $c$-axis of the crystal ($\theta = 0^\circ$) is rotated toward the field direction, higher magnetic fields are needed to suppress long-range order.}
\label{fig:ehrenfest}
\end{figure}

The experimentally observed jump $\Delta k \approx$  6 J/rad$^2$/mol in this configuration (Figure \ref{fig:jump}b) can be directly compared to heat capacity measurements under magnetic field. $\der{T_c}{θ}$ can be estimated from the angle dependence of the resonant torsion of RuCl$_3$ at fixed temperature and magnetic field.
One such scan at T = 1.3 K and H = 17.5 T (Figure \ref{fig:ehrenfest}a) shows a pronounced anomaly at the phase boundary of the long-range ordered state. Entry $\it into$ the ordered state is marked by a jump $\it down$ at the phase boundary, as required by Equation \ref{eq:ehrenfest}. Such measurements at various fixed magnetic fields allow to map out the phase boundary of the antiferromagnetically ordered state (Figure \ref{fig:ehrenfest}b). The derivative $\der{T_c}{ \theta}_H=\der{T_c}{H}_\theta \der{H_c}{\theta}_T$ at T = 1.3 K, H = 10 T, and $\theta$ = 102$^\circ$ can be estimated as $\der{H_c}{\theta}_T \approx 2.8$ T/rad and $\der{T_c}{H}_\theta \approx 25$ K/T. The heat capacity jump at the antiferromagnetic transition at T = 1.3 K has been reported as $ΔC/T_c \sim$ 1.7 mJ/mol/K$^2$ \cite{heatcapacity}. Thus the right hand side of Eq. \ref{eq:ehrenfest} gives approximately 8 J/rad$^2$/mol, in agreement with the size of the measured jump $\Delta k$ of 6 J/rad$^2$/mol found above. This quantitative agreement is remarkable, especially given the uncertainties of the derivatives due to the complex shape of the phase boundary.

Resonant torsion magnetometry is a highly sensitive, quantitative and simple method to probe the magnetic anisotropy of the smallest crystallites of quantum matter. The magnetotropic coefficient provides valuable information about anisotropic materials, complementing the magnetic torque. First, direct measurement of $k$ clearly signals second-order phase transitions by discontinuous jumps that can be related to anomalies in other thermodynamic measurements. Resonant torsion magnetometry thus serves as an alternative tool for measuring magnetic phase transitions in environments that are challenging for other thermodynamic techniques. Second, measurements of this coefficient allows direct access to the magnetic anisotropy when magnetic field is aligned $\it along$ the principal magnetic axes -- a blindspot for conventional torque magnetometry. Finally, the ability to measure shifts in the resonant frequency of lever vibrations much more precisely than the amplitude of lever deflections results in better than part per 100 million sensitivity and the opportunity to measure sub-nanogram sized samples. In the area of materials discovery, microscopic crystallites commonly occur and resonant torsion magnetometry provides a new route to conveniently determine the anisotropy of new quantum materials.

\section{Acknowledgments}
The authors would like to thank the electronics workshop at the Max Planck Institute for Chemical Physics of Solids, particularly Dominic Hibsch, Wolfgang Geyer, and Torsten Breitenborn. We also thank Terunobu Akiyama for helpful discussions. Synthesis and characterization of the NbP single crystals was performed at Los Alamos National Laboratory under the auspices of the US Department of Energy, Office of Basic Energy Sciences, Division of Materials Sciences and Engineering. The portion of this work completed at the National High Magnetic Field Laboratory is supported through the National Science Foundation Cooperative Agreement numbers DMR-1157490 and DMR-1644779, The United States Department of Energy, and the State of Florida. M.D.B. acknowledges studentship funding from EPSRC under grant no. EP/I007002/1. RDM acknowledges support from LANL LDRD-DR 20160085 topology and strong correlations.

\bibliographystyle{apsrev4-1}
\bibliography{refs}

\clearpage

\section{Methods}

Our measurements were done in a 16 T superconducting magnet and a 35 T resistive Bitter magnet.  Approximately 10 mbar of exchange gas was maintained in the sample chamber during these measurements. Under these conditions, we found that an optimal drive voltage of 10 mV provides good sensitivity without causing the lever to ring. Frequency scans with an output voltage as high as 50 mV show no shift in the observed resonant frequency of the cantilever, confirming that the lever responds linearly to the deflection.

\begin{figure}[htbp]
\centering
\includegraphics[width=1.01\linewidth, trim=4.2cm 5cm 16.5cm 0cm, clip=true]{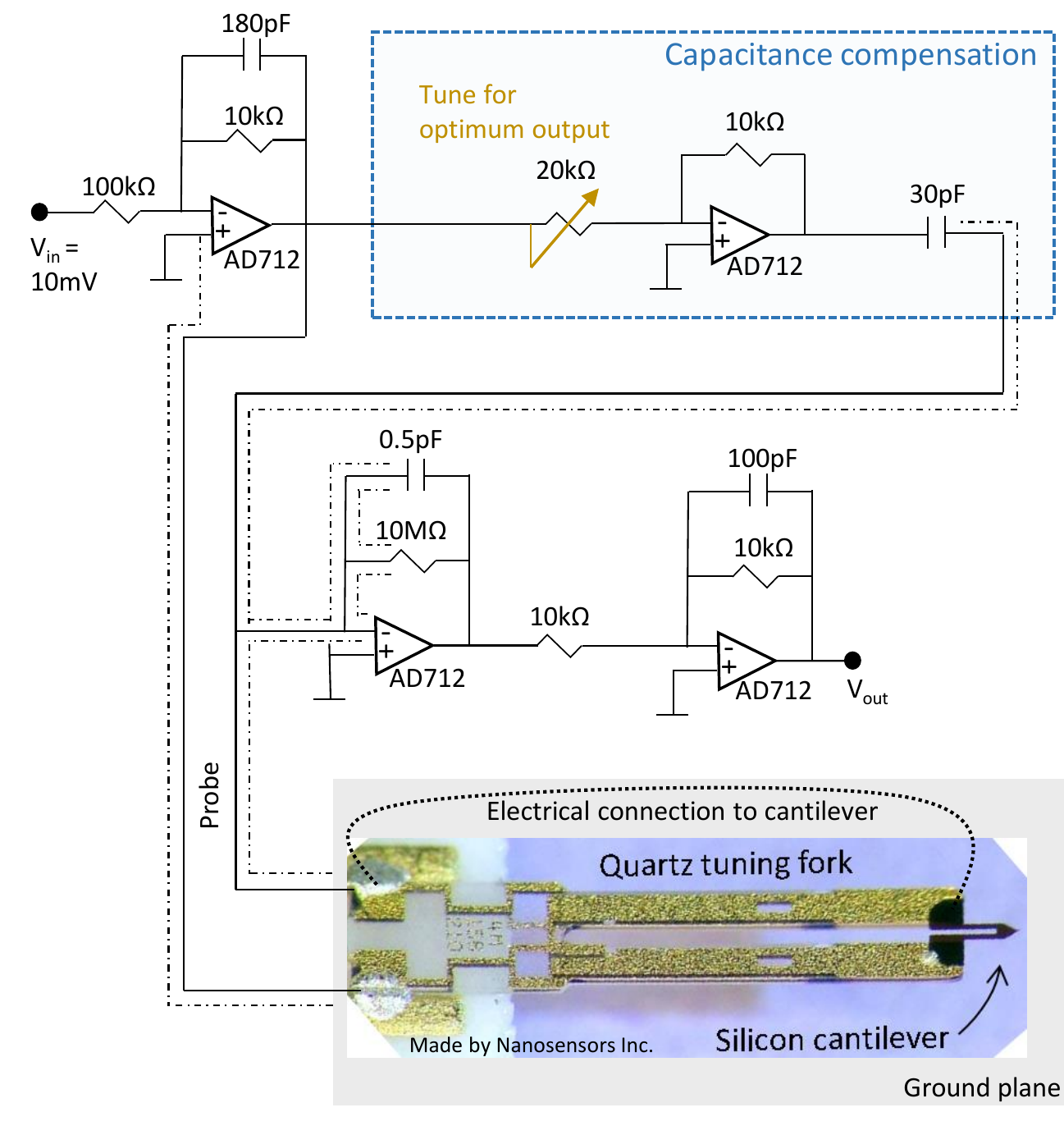}
\rule{27em}{0.5pt}
\caption[Circuit diagram]{\scriptsize Schematic diagram of the electrical circuit including capacitance compensation, amplification, and placement of the A-Probe \cite{akiyama} in the circuit.}
\label{fig:circuit}
\end{figure}

The experimental setup requires a method for tracking the frequency of the mechanical resonance of the cantilever as a function of temperature and magnetic field. The motion of the cantilever couples to the tuning fork, which has two contacts used to measure the electrical resonance of the parallel LC circuit. The pickup signal is lockin-detected at the drive frequency. Due to the internal capacitance of the tuning fork and cabling, the voltage induced on the pickup contact is non-zero and has a finite phase with respect to the drive signal, even when the vibration amplitude is negligible away from the resonance. The finite background impedance, which also changes as a function of temperature, may inhibit the magnitude of the sharp phase change expected on resonance. A large phase change across the resonance is required for successful tracking with the phase-locked loop (PLL). To correct for the background phase, we incorporate a capacitance compensation circuit based on the one recommended by Nanosensors Inc. (Figure \ref{fig:circuit}) \cite{akiyama, capcircuit}.

In order to precisely measure the small piezoelectric current due to the mechanical motion of the cantilever at resonance, the parasitic capacitance between the gold contacts on the tuning fork must be compensated in the circuit. Our experimental setup is shown in Figure \ref{fig:circuit}. Here, the output from the capacitance compensation line is inverted and summed with the measured response of the cantilever at resonance to detect $\it only$ the piezoelectric current due to the lever motion. At zero magnetic field, the background capacitance is nulled with the potentiometer in order to obtain a large 180$^\circ$ phase shift on resonance.  To reduce noise, the braided shield of a low-capacitance twisted pair directly connects the positive input of the current-voltage converter to a ground plane directly below the vibrating lever (Figure \ref{fig:circuit}). The signal is then amplified to follow the resonant frequency with the PLL as it changes with magnetic field. We used the (PLL/PID) option of the Zurich Instruments mid-frequency lock-in (MFLI) amplifier with analog to digital to analog conversion. As a function of temperature, large changes in the background impedance (due to thermal contraction of the wiring, the inductance in electrical components, etc.) are observed. Thus, we use a custom program to adaptively follow the resonance, which maintains that the in-phase and quadrature components of the frequency scan around the resonance are centered at zero \cite{arkady, rus}.

$K$ denotes an $\it effective$ bending stiffness of the lever. The actual shape of the cantilever can be described in the thin-plate approximation \cite{landau} by $ζ(z,t)$, where $ζ$ is the displacement of the lever at a distance $z$ from the point of attachment. In the thin-plate approximation, the shape of $ζ(z,t)$ can be determined from the energy functional $E = (1/2) ∫_0^L dz ρ(z)A(z)(dζ(z,t)/dt)^2 + (1/2)  ∫_0^L  dz E(z) I_y(z) (\Delta ζ(z,t))^2$, where $A(z)$ is the cross-sectional area of the lever, $I_y(z)$ is the moment of inertia of the cross section of the lever, $L$ is the length and $ρ(z)$ is the density. This form of energy allows a straightforward derivation of the boundary conditions for a cantilever of non-uniform crossection. With this, $∆θ$ is defined as the rotation angle at the tip of the lever $∆θ(t) = (dζ(z)/dz)_{z=L}$. Thus, $K$ and $I$ in Eq. \ref{eq:one} must be found from the detailed solution for the shape of the resonance mode. Not only do $K$ and $I$ depend on the shape of the lever, both have different values when different resonant modes are considered in Eq. \ref{eq:one}. For example, the effective bending stiffness for the lowest oscillating mode (without nodes) of a uniform cross-section cantilever is $K_0=1.63 (EI_y/L)$. The effective moment of inertia for the same mode is $I_0 =  0.13 ρAL^3$.

\end{document}